\documentclass[
aps,
prb,
notitlepage,
twocolumn,
superscriptaddress,
longbibliography,
nofootinbib
]{revtex4-2}
\pdfoutput=1
\usepackage[utf8]{inputenc}
\usepackage{amsmath,amsthm,amssymb,amsfonts}
\usepackage{scalerel}
\usepackage{mathtools}
\usepackage{graphicx}
\usepackage{subfigure}
\usepackage[normalem]{ulem}
\usepackage[colorlinks = true,
            linkcolor = blue,
            urlcolor  = brown,
            citecolor = red,
            anchorcolor = blue]{hyperref}
\usepackage{mathrsfs}
\usepackage{bbold}
\usepackage{units}
\allowdisplaybreaks
\usepackage{bm}
\usepackage{braket}
\usepackage{makecell}
\usepackage[nameinlink]{cleveref} 
\usepackage{upgreek}
\usepackage{blindtext}
\usepackage{verbatim}
\usepackage{algorithm}
\usepackage[noend]{algpseudocode}
\usepackage{balance}
\makeatother

\usepackage[dvipsnames]{xcolor}
\usepackage{float}
\usepackage[bb=boondox]{mathalfa}
\usepackage{array}
\usepackage{multirow}
\usepackage{tabularx}
\usepackage{float}
\usepackage{dcolumn}
\usepackage{slashed}
\usepackage{braket}
\usepackage{verbatim}
\usepackage{multirow}
\usepackage{resizegather}
\usepackage{titlesec}
\usepackage[toc,page]{appendix}
\usepackage[export]{adjustbox}
\definecolor{teal}{RGB}{0, 158, 115} 
\definecolor{morange}{RGB}{255, 127, 0}

\definecolor{titleblue}{RGB}{37,96,167}

\begin{document}

\title{\textcolor{titleblue}{Full Eigenstate Thermalization in Integrable Spin Systems}}

\author{Tanay Pathak\,\,\href{https://orcid.org/0000-0003-0419-2583}
{\includegraphics[scale=0.05]{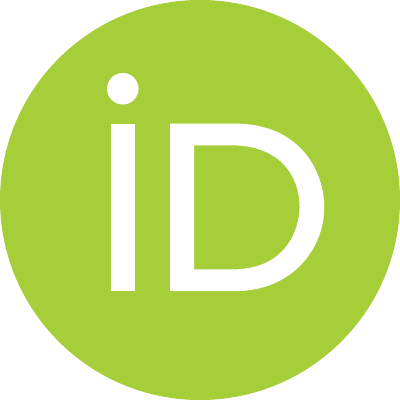}}}
\email{pathak.tanay.4s@kyoto-u.ac.jp}
\affiliation{Department of Physics, Kyoto University, Kitashirakawa Oiwakecho, Sakyo-ku, Kyoto 606-8502, Japan}

\begin{abstract}
The Eigenstate Thermalization Hypothesis (ETH) is a standard tool to understand the thermalization properties of an isolated quantum system. Its generalization to higher order correlations of matrix elements of local operators, dubbed the full ETH, relates the ETH free cumulants to the corresponding thermal free cumulants in the thermodynamic limit. In this work, we numerically test the full ETH predictions using exact diagonalization of two integrable spin models: the Ising and the XXZ Heisenberg models. The differences from the behavior predicted by full ETH in chaotic systems are highlighted and contrasted along the way. We further study the out-of-time-ordered correlator (OTOC) through its approximate decomposition into contributions involving the second- and fourth-order ETH free cumulants. We find that, although the fourth-order contribution governs the late-time behavior of the OTOC in these integrable models, its dynamics differs significantly from that found in nonintegrable systems.
\end{abstract}

\maketitle

\section{Introduction} The emergence of statistical mechanics in quantum many body systems is an important and challenging problem. Currently, it is mostly understood using the eigenstate thermalization hypothesis (ETH) \cite{Deutsch:1991msp,Srednicki:1994mfb,Deutsch:2025pmk}. ETH states that a local observables $\hat{O}$,  written in the energy eigenbasis, are (pseudo)random banded matrices whose statistical properties are smooth thermodynamics functions. ETH has been well verified in many nonintegrable many body systems \cite{PhysRevE.60.3949,Rigol:2007juv,PhysRevLett.105.250401,RevModPhys.83.863,PhysRevE.87.012125,DAlessio:2015qtq,PhysRevLett.117.170404,PhysRevB.93.134201}. The standard ETH however can only describe correlation amongst two matrix elements relevant to describe upto two point functions. However, it is also important to consider multi-point correlators from the perspective of quantum chaos and scrambling such as the out-of-time-ordered correlator (OTOC) \cite{larkin1969quasiclassical,Maldacena:2015waa,Hosur:2015ylk,Roberts:2016hpo,Hashimoto:2017oit,Garcia-Mata:2023}. The importance of correlations amongst the matrix elements was already highlighted in \cite{PROSEN1994115} and these correlation have also been subject of recent interests \cite{Foini:2018sdb,PhysRevLett.123.260601,PhysRevLett.122.220601,PhysRevLett.123.230606,PhysRevE.102.042127,PhysRevE.104.034120}. Recently ETH has been generalized to take into account the correlations among multiple matrix elements, dubbed as \emph{ full ETH} \cite{Foini:2018sdb,Pappalardi:2022aaz}. Further relation of full ETH with \emph{free probability} (non commutative probability) \cite{danfreeprob,Speicher2003,nica2006lectures,speicher2017free,mingo2017free} have also been found, which has further application in many body physics and in the understanding of quantum chaos \cite{Pappalardi:2023nsj,PhysRevX.13.011045, Jindal:2024zcg,PhysRevB.111.014311,Jahnke:2025exd,Camargo:2025zxr,Fritzsch:2025arx,Dowling:2025cxr,Fritzsch:2025ban,PhysRevB.111.054303}. 

ETH applies to nonintegrable systems, for which the expectation values of observables are believed to reach thermal (microcanonical ensemble) values at late times. There have also been considerable studies of ETH in integrable systems where it is expected to not hold \cite{PhysRevE.87.012118,PhysRevB.91.155123,PhysRevE.91.012144,PhysRevE.102.062113,PhysRevLett.113.117202,Vidmar_2016,PhysRevLett.125.070605,PhysRevX.14.031048,Rottoli:2025pzr}. It is then natural to question the validity of \emph{full} ETH conditions in integrable systems. In this work we investigate the question: \emph{What is the fate of the full ETH conditions in integrable systems and how do they differ from nonintegrable cases?} We answer this question by numerically studying integrable spin model viz the Ising and the XXZ models. Specifically, we study \emph{full} ETH properties of these model by studying a four point correlator, the OTOC.

\section{Full ETH.} The standard ETH ansatz \cite{Deutsch:1991msp,Deutsch:2025pmk,Srednicki:1994mfb}, for an observable $\mathcal{O}$, in the energy eigenbasis of the Hamiltonian($H \ket{n}= E_{n}\ket{n}$) is given as 
\begin{equation}
\bra{m}\hat{O} \ket{n}= O_{mn}= O(E^{+}) \delta_{mn}+ R_{mn} e^{-\frac{S(E^+)}{2}} F(E^{+}, E_{m}-E_{n}),
\end{equation}
where $E^{+}= (E_{m} + E_{n})/2$ is the mean energy and, $R_{mn}$ is a random Gaussian number of zero mean and unit variance, $S(E^+)$ is the thermodynamic entropy and $F(E^{+}, E_{m}-E_{n})$ is a function of order 1. This ansatz however do not account for the higher order correlations found in the matrix elements. Based on typicality argument, the ETH ansatz was thus generalized to the product of $q$-- distinct matrix elements \cite{Foini:2018sdb,Pappalardi:2022aaz}, for an operator $O$ written in the eigenbasis of a given Hamiltonian $H$ and is given by 
\begin{equation}\label{eqn:fulleth}
\overline{O_{i_{1} i_{2}}O_{i_{2} i_{3}} \cdots O_{i_{q}i_{1}}} =  e^{-(q-1)S(E^{+})} F^{(q)}_{e^+}(\omega)
\end{equation}
where the \emph{overbar} on the left hand side denotes averaging, $e^{-S(E^{+})}$ is the mean level spacing at average energy $E^{+}= \frac{E_{1}+\cdots + E_{q}}{q}$, $\omega= (\omega_{i_{1}i_{2}}, \cdots, \omega_{i_{q-1}i_{q}})$; $\omega_{jk}= E_{j}-E_{k}$ are the energy differences and $F^{(q)}_{e^+}(\omega)$ is a smooth order one function of the energy density $e^+= E^+/N$ and $\omega$. Eq. \eqref{eqn:fulleth} is referred as \emph{full ETH}. The thermal multi-time correlation functions are related to the thermal free cumulants through the general moment–free-cumulant relation as follows
\begin{equation}\label{eqn:multipoint}
\braket{O(t_{1})O(t_{2}) \cdots O(t_{q})}_{\beta} = \sum_{\pi \in N C(q)} \kappa_{\pi}^{\beta}(t_{1}, t_{2}, \cdots, t_{q}), 
\end{equation}
where $\braket{\bullet}_{\beta}= \frac{1}{Z} \text{Tr}(e^{-\beta H} \bullet)$ is the thermal average at inverse temperature $\beta$. $\kappa_{\pi}^{\beta}$ denotes the product of thermal free cumulants $k_{n}^{\beta}$, corresponding to each block of the non-crossing partition, $\pi$ \cite{noncross}. The full ETH ansatz, given by Eq. \eqref{eqn:fulleth}, then implies that the thermal free cumulants \emph{may} be approximated by ETH free cumulants for computing  multi time correlations as follows
\begin{equation}\label{eqn:kq}
    k_{q}^{\beta}(\vec{t}) \simeq k_{q}^{\text{ETH}}(\vec{t})= \frac{1}{Z} \sum_{i_{1} \neq i_{2}\neq \cdots \neq i_{q}} e^{-\beta E_{i_1}} O(t_{1})_{i_{1}i_{2}}O(t_{2})_{i_{2}i_{3}} \cdots O(t_{q})_{i_{q}i_{1}}.
\end{equation}
where the relation $k_{q}^{\beta}(\vec{t}) \simeq k_{q}^{\text{ETH}}$ is expected to hold only in the thermodynamic limit.
It is known that the usual ETH gives $\braket{O(t)O(0)}_{\beta}\simeq [k^{\text{ETH}}_{1}]^{2}+ k^{\text{ETH}}_{2}(t)$ \cite{DAlessio:2015qtq}, for the two-point function. Eq. \eqref{eqn:kq} then generalizes this relation to higher point correlators. 
The contribution of the free cumulants can be then represented in a diagrammatic manner with two types of diagrams: crossing and non-crossing, similar to the diagrammatic formulation of classic moment-cumulant formula differing from it earliest at the fourth order \cite{Pappalardi:2022aaz}. It was shown in \cite{Pappalardi:2022aaz} that the ETH ansatz \eqref{eqn:fulleth} implies the following:
(i) the crossing contribution are exponentially subleading. For instance, considering the relevant contribution to the four point function would imply   
\begin{equation}\label{eqn:fulleth1}
   \text{cross}(t)= \frac{1}{D} \sum_{i \neq j}e^{i t \omega_{ij}}|O_{ij}|^4= \mathcal{O}(D^{-a}),\, a>0, 
\end{equation}
and (ii) the non-crossing diagram or cactus diagram (corresponding to noncrossing partition in Eq. \eqref{eqn:multipoint}) factorize. For instance
\begin{equation}\label{eqn:fulleth2}
\text{cac}(t_{1},t_{2})= \frac{1}{D}\sum_{i \neq j \neq k} e^{i \omega_{ij}t_{1}+ i \omega_{jk} t_{2}} |O_{ij}|^{2} |O_{jk}|^{2} \approx k^{\text{ETH}}_{2}(t_{1}) k^{\text{ETH}}_{2}(t_{2}),
\end{equation}
for generic times $t_{1}$ and $t_{2}$, $k^{\text{ETH}}_{2}(t)$ is given by Eq. \eqref{eqn:kq} and $D$ is the Hilbert space dimension. The conditions given by Eq. \eqref{eqn:fulleth1} and \eqref{eqn:fulleth2} hold both in time as well as frequency domain. 

\section{Models} 
We focus our study on the following two paradigmatic integrable spin models.

1. The Hamiltonian of the Ising model with transverse and longitudinal fields is given by \cite{PhysRevLett.106.050405,PhysRevLett.111.127205, PhysRevE.91.062128,Roberts:2014isa, PhysRevB.101.174313,Rodriguez-Nieva:2023err} 
\begin{equation}
    H = \sum_{i=1}^{L-1} J \sigma_{i}^{z} \sigma_{i+1}^{z} + \sum_{i=1}^{L} h_{x}\, \sigma_{i}^{x}  + \sum_{i=1}^{L} h_{z}\, \sigma_{i}^{z},
\end{equation}
where $\sigma_{i}^{a}$ $a \in \{x,y,z\}$ denotes a Pauli matrix at site $i$ with $a \in \{x, y, z\}$. With open boundary condition, considered here, the model has space reflection symmetry. In our study we consider the integrable regime parameters as: $J= 1$, $h_{x}= -1$, $h_{z}= 0$ and also provide results of chaotic case, with parameters: $J= 1$, $h_{x}= -1.05$, $h_{z}= 0.5$ (see \cite{Pappalardi:2023nsj} and also Appendix \ref{appendix:isingchao}). We focus on the following traceless local operator in the bulk 
\begin{equation}\label{eqn:oising}
   \hat{O}= \sigma_{L/2}^{z}. 
\end{equation}

2. The Heisenberg XXZ spin chain is given as
\begin{equation}
    H_{\text{XXZ}} = \sum_{i=1}^{L-1}J (\sigma_{i}^{x} \sigma_{i+1}^{x}+\sigma_{i}^{y} \sigma_{i+1}^{y} + \Delta \sigma_{i}^{z} \sigma_{i+1}^{z}),
\end{equation}
where $\Delta$ is the anisotropy parameter and we choose $J=1$ for our analysis. With open boundary condition the model has space reflection symmetry and rotation symmetry (which conserves the total magnetization). We explicitly break the space reflection symmetry by adding a small impurity at the first site, $h_{1}\sigma_{1}^{z}$, with field strength $h_{1}= 0.001$, which does not break the integrability \cite{Alcaraz:1987uk, Santos_2004}. It has been shown that even a small defect away from the edges in the chain can lead to level repulsion and random matrix type statistics of the spectrum in the system \cite{Alcaraz:1987uk, Santos_2004,PhysRevE.84.016206,PhysRevE.89.062110,Torres_Herrera_2015,PhysRevB.80.125118,PhysRevB.98.235128}. We focus on $\Delta = 0.55$, easy-plane regime \cite{PhysRevLett.106.217206, PhysRevLett.111.057203, PhysRevB.98.235128, RevModPhys.93.025003}, and $\Delta = 1.1$, the easy-axis regime. We thus consider the Hamiltonian
\begin{equation}
   H =  H_{\text{XXZ}} + h_{1}\sigma_{1}^{z}+ d\,\sigma^{z}_{L/2},
\end{equation}
where $d$ is of the order of $J$. We choose $d= 0$ for integrable case and $d= 1$ for nonintegrable case. 
For XXZ model we study following two local operators
\begin{equation}\label{eqn:opxxz}
  \hat{O}_{1} = \sigma_{L/4}^{z} , \quad \hat{O}_{2} = \sigma_{3L/4}^{x}\sigma_{3L/4+1}^{x} +\sigma_{3L/4}^{y}\sigma_{3L/4+1}^{y},
\end{equation}
far away from the edge and the bulk impurity. With this choice of bulk operators the OTOC dynamics is not qualitatively affected by the boundary perturbation that breaks the reflection-symmetry. We perform full diagonalisation of spin chain upto $L= 18$ spins, focusing only on the zero magnetization sector. 

\section{Results}
\begin{figure}[htbp]
\centering
\includegraphics[width= \linewidth]{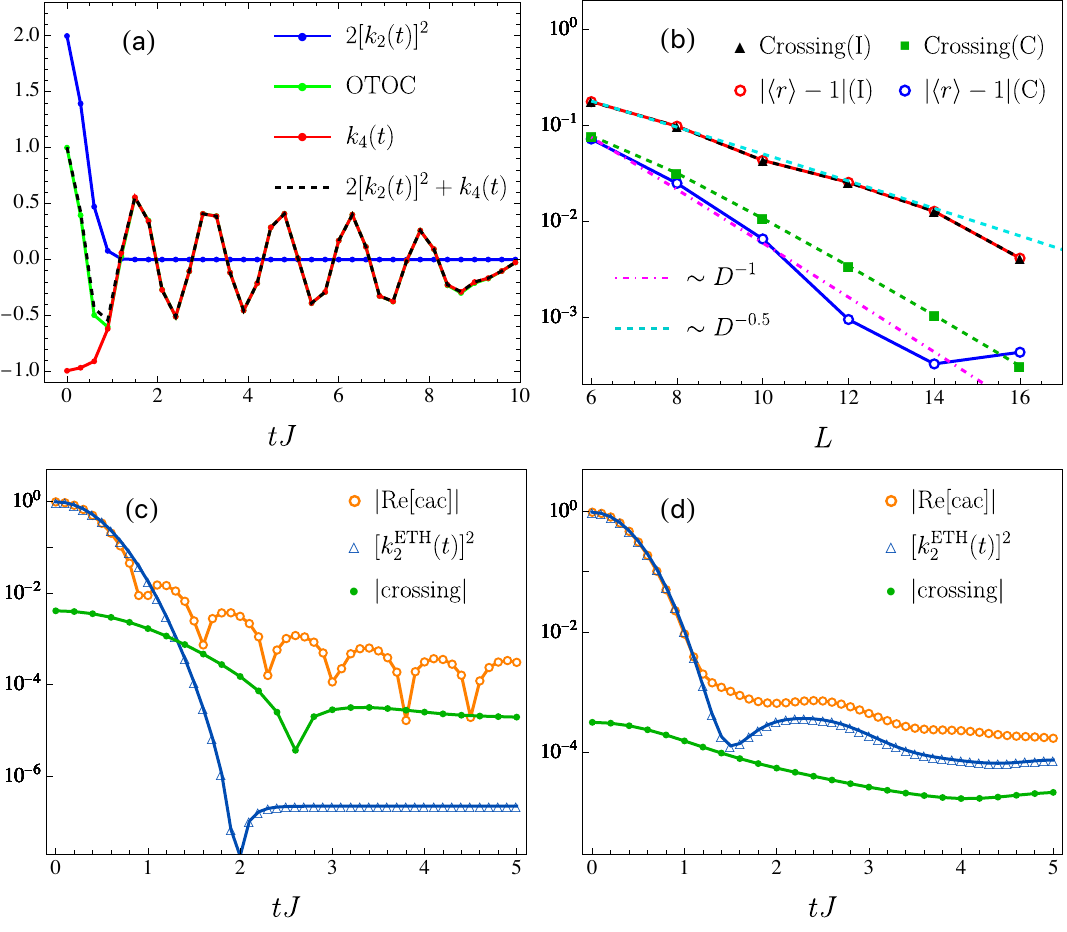}
\caption{(a) The time (measured in unit of $tJ$) evolution OTOC compared with $2[k^{\text{ETH}}_{2}(t)]^{2}$ and $k^{\text{ETH}}_{4}(t)$ along with the sum of the two, for integrable Ising model (denoted by I). (b) Variation of $|\langle r \rangle -1|$ and crossing contribution for integrable and chaotic (denoted by C) Ising model. Also shown are the $D^{-1}$ scaling (magenta line) to be compared with the chaotic case and the $D^{-0.5}$ scaling (cyan line) to be compared with the integrable case. Factorization of cactus diagram in time domain and crossing contribution as a function of time (c) for integrable Ising chain and (d) for chaotic Ising chain. The results in (a), (c) and (d) are for spin chains of length $L=16$. The solid lines are guide to the eyes.}
\label{fig:isingfulleth}
\end{figure}

We first examine the relation between the OTOC and the free cumulants obtained from the full ETH ansatz, at infinite temperature i.e. $\beta=0$. For a traceless operator, validity of full ETH gives the following relation \cite{Foini:2018sdb, Pappalardi:2022aaz}
\begin{equation} \label{eqn:otocfulleth}
\text{OTOC}(t)\simeq 2[k^{\text{ETH}}_{2}(t)]^{2} + k^{\text{ETH}}_{4}(t),
\end{equation}
where $k^{\text{ETH}}_{2}(t)$ and $k^{\text{ETH}}_{4}(t)$ are the free cumulants defined in Eq. \eqref{eqn:kq}, and $\text{OTOC}(t)=\braket{O(t)O(0) O(t)O(0)}_{\beta=0}$ follows from Eq. \eqref{eqn:multipoint} with $t_1=t_3=t$ and $t_2=t_4=0$. This relation was previously verified numerically for local operators in chaotic lattice systems at both early and late times \cite{Pappalardi:2023nsj} (see also Appendix \ref{appendix:isingchao}). Using the operator $\hat{O}$ in Eq. \eqref{eqn:oising}, we find that Eq. \eqref{eqn:otocfulleth} also holds in the integrable case, as shown in Fig. \eqref{fig:isingfulleth}(a). Moreover, as in chaotic systems, the late-time behavior of the OTOC is governed by $k^{\text{ETH}}_{4}(t)$. The effect of integrability appears in the dynamics: compared with the chaotic case, the OTOC and $k^{\text{ETH}}_{4}(t)$ decay more slowly and show persistent oscillations over the accessible time window. Here, by \emph{late times} we mean the post-transient regime in the finite-size numerics i.e. $tJ\gtrsim 1$, where we observe that the OTOC and $k^{\text{ETH}}_{4}(t)$ begin to follow similar behavior.

Eq. \eqref{eqn:otocfulleth} is expected to hold when crossing diagrams are suppressed and cactus diagrams factorize. Since $\text{cac}(t)$ has a small finite-size imaginary part that decreases with $L$ and is expected to vanish as $L\rightarrow\infty$, we compare $|\mathrm{Re}[\text{cac}(t)]|$ with $[k^{\text{ETH}}_2(t)]^2$ in Figs. \eqref{fig:isingfulleth}(c,d). In both the integrable and chaotic chains, factorization holds at early times and breaks down at later times. The deviations are larger in the integrable case. However, this late-time breakdown has little visible effect on Eq. \eqref{eqn:otocfulleth}: for $tJ>1$, $[k^{\text{ETH}}_2(t)]^2\lesssim 10^{-2}$, while $k^{\text{ETH}}_4(t)$ already gives the dominant contribution to the OTOC. The crossing contribution is small for both integrable and chaotic chains over the full time window considered.

To quantify finite-size effects in the factorization, we study \cite{Pappalardi:2023nsj}
\begin{equation}
    \braket{r}= \frac{\text{cac}(0,0)}{[k^{\text{ETH}}_2(0)]^2}.
\end{equation}
Full ETH predicts $\braket{r}\rightarrow 1$ as $L\rightarrow\infty$. Figure \eqref{fig:isingfulleth}(b) shows that $|\braket{r}-1|$ decreases with system size in both integrable and chaotic Ising chains. The decay is slower in the integrable case; for $L\geq 12$, the chaotic values are at least one order of magnitude smaller than the integrable ones. We also examine the crossing contribution at $t=0$, where it is maximal. The integrable system show a slower scaling, roughly $\sim D^{-0.5}$, whereas the chaotic systems results show faster $\sim D^{-1}$ decay. Similar qualitative behavior is obtained for other local operators (see Appendix \ref{appendix:othero}) as well.

\begin{figure}
\centering
\includegraphics[width= \linewidth]{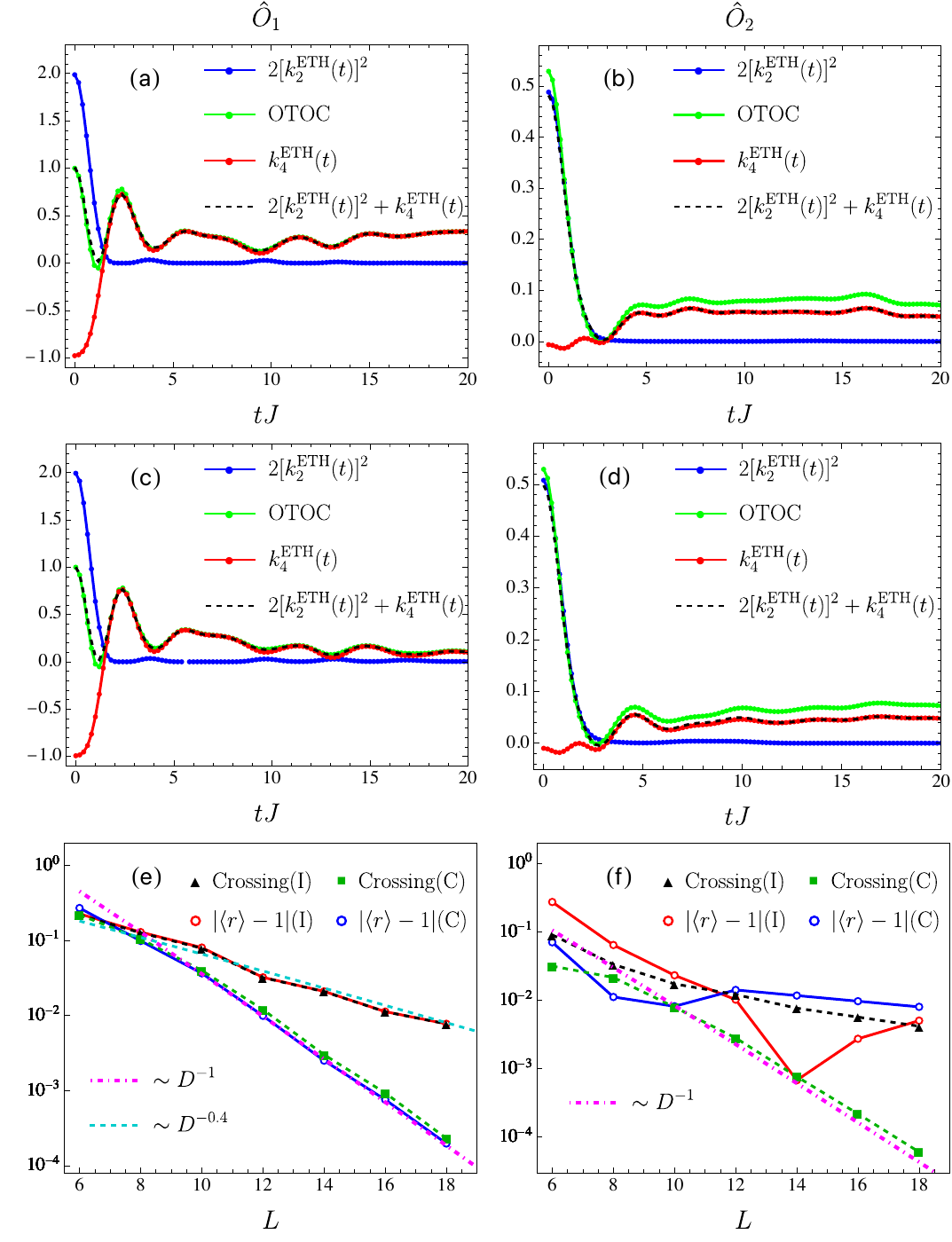}
\caption{The time evolution OTOC, compared with $2[k^{\text{ETH}}_{2}(t)]^{2}$ and $k^{\text{ETH}}_{4}(t)$ for XXZ model with $L=18$ and $\Delta= 0.55$(easy-plane). Also shown is the sum of the two free-cumulant contributions. The results for $\hat{O}_1$ are shown in (a)for the integrable case (c) for the chaotic case while those for $\hat{O}_2$ are shown in (b) integrable case (d) chaotic case. Scaling of $|\braket{r}-1|$, and crossing contribution with system size $L$ (e) for operator $\hat{O}_{1}$ (f) for operator $\hat{O}_{2}$. The solid lines are guide to the eyes.}
\label{fig:xxzfullethballistic}
\end{figure}

Next, we consider the XXZ model in the easy-plane regime, $\Delta=0.55$. For the operator $\hat{O}_1$, Figs. \eqref{fig:xxzfullethballistic}(a,c) show the OTOC in the integrable $(d=0)$ and chaotic $(d=1)$ cases, together with $2[k^{\text{ETH}}_2(t)]^2$ and $k^{\text{ETH}}_4(t)$. Equation \eqref{eqn:otocfulleth} holds well in both cases, and the late-time OTOC is again controlled mainly by $k^{\text{ETH}}_4(t)$. The difference between the two cases appears primarily in the late-time saturation value, which is lower in the nonintegrable chain. For the second operator, $\hat{O}_2$, shown in Figs. \eqref{fig:xxzfullethballistic}(b,d), the OTOC-cumulant relation still remains valid and $k^{\text{ETH}}_4(t)$ still captures the late-time dynamics of the OTOC. However, the late-time saturation values are close in the integrable and chaotic cases, approximately $0.069$ and $0.058$, respectively, for $tJ\geq 15$. Thus, for this observable, the distinction between integrable and chaotic XXZ dynamics is much weaker than in the Ising model.

\begin{figure}
\centering
\includegraphics[width= \linewidth]{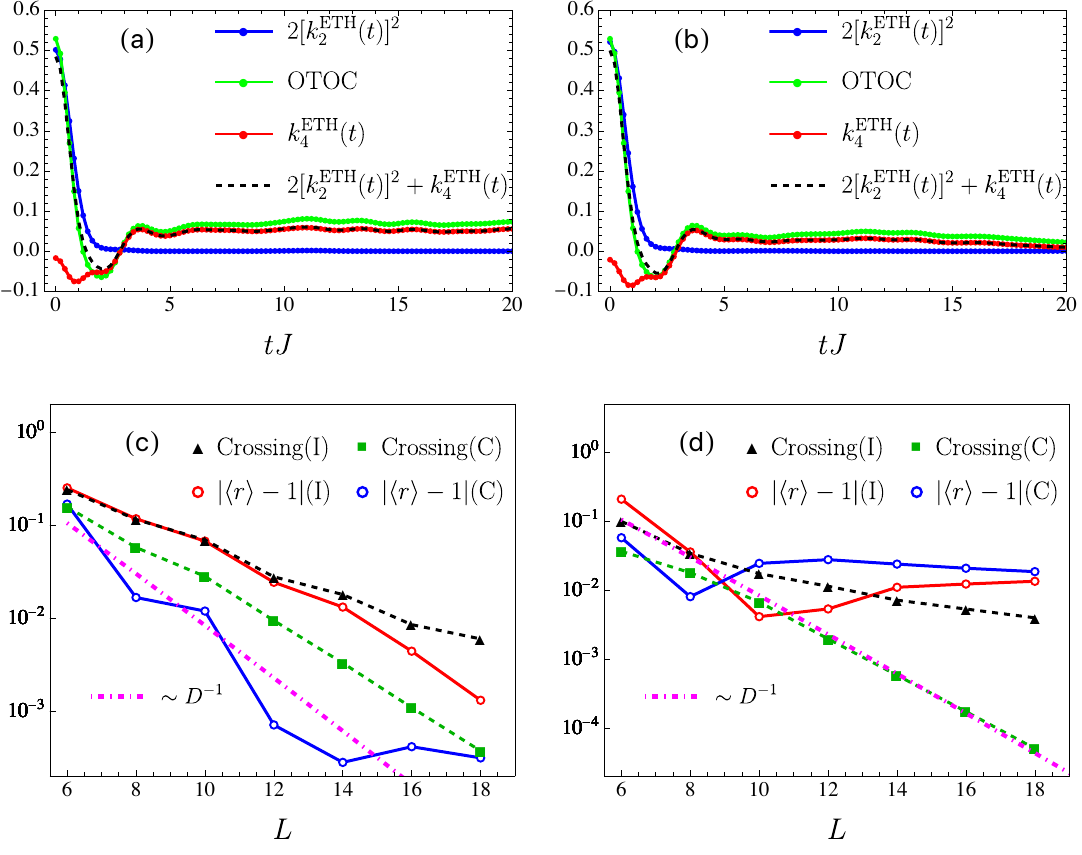}
\caption{The time evolution OTOC, compared with $2[k^{\text{ETH}}_{2}(t)]^{2}$ and $k^{\text{ETH}}_{4}(t)$ for XXZ model; $L=18$ and $\Delta= 1.1$ (easy-axis). Also shown is the sum of the two free cumulant contributions.(a) For $d=0$ and operator $\hat{O}_2$. (b) For $d=1 $ and operator $\hat{O}_2$. Scaling of $|\braket{r}-1|$ and crossing contribution, with system size $L$, (c) for operator $\hat{O}_{1}$ and (d) for operator $\hat{O}_{2}$}.
\label{fig:xxzfullethdiffusive}
\end{figure}

We now test the two full-ETH conditions in Eqs. \eqref{eqn:fulleth1} and \eqref{eqn:fulleth2}. For $\hat{O}_1$, Fig. \eqref{fig:xxzfullethballistic}(e) shows behavior similar to the Ising chain: the crossing contribution decays with different indicative scalings in the integrable and chaotic cases, approximately $\sim D^{-0.4}$ and $\sim D^{-1}$, respectively. For $\hat{O}_2$, however, Fig. \eqref{fig:xxzfullethballistic}(f) shows that $|\braket{r}-1|$ does not clearly distinguish the integrable and chaotic cases over the accessible system sizes. The crossing contribution still separates the two regimes, but the cactus-factorization correction in the chaotic chain is not visibly suppressed as $\mathcal{O}(D^{-1})$. This indicates that, at the sizes studied here, certain observables in the locally perturbed chaotic XXZ chain retain integrable-like features. This behavior is reminiscent of Ref. \cite{PhysRevLett.125.070605}, where the diagonal, standard ETH in a locally perturbed model was found to follow the microcanonical prediction of the integrable model.

\begin{figure}
\centering
\includegraphics[width= \linewidth]{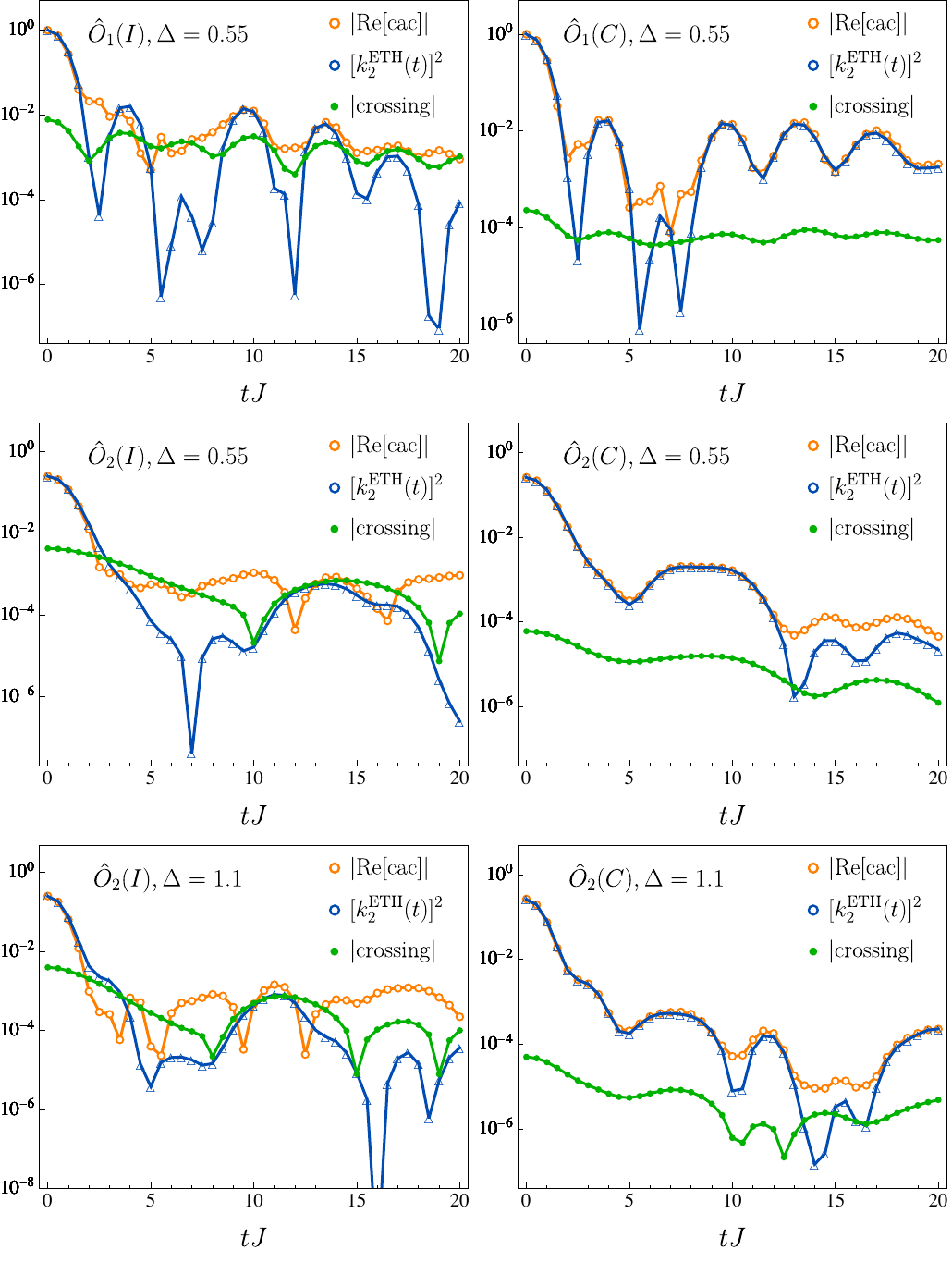}
\caption{Factorization of cactus diagram in the time domain for XXZ model; $L= 18$. (First row) Operator $\hat{O}_{1}$, for integrable (denoted by I) and chaotic (denoted by C) case respectively. (Second row) Operator $\hat{O}_{2}$, and integrable and chaotic case respectively for $\Delta=0.55$.. (Third row) Operator $\hat{O}_{2}$, and integrable and chaotic case respectively for $\Delta=1.1$. Also shown, in each plot, is the corresponding crossing contribution as a function of time (green markers).}
\label{fig:xxzfactorcross}
\end{figure}

We repeat the analysis for the easy-axis regime, $\Delta=1.1$. The results for $\hat{O}_2$ are shown in Figs. \eqref{fig:xxzfullethdiffusive}(a,b), while the corresponding results for $\hat{O}_1$ are given in Appendix \ref{appendix:xxzothero}. Compared with the easy-plane regime, the chaotic chain shows better agreement with Eq. \eqref{eqn:otocfulleth} at early times and saturates to a smaller late-time value. The integrable chain remains qualitatively similar. The finite-size behavior of $|\braket{r}-1|$ and of the crossing contribution, shown in Figs. \eqref{fig:xxzfullethdiffusive}(c,d), is also qualitatively similar to the easy-plane regime. This also suggests that the observed scaling behavior is not strongly affected by the spin transport properties in the two regimes namely ballistic transport for easy-plane regime ($\Delta <1$) and diffusive transport for easy-axis regime ($\Delta >1$).

Finally, we examine cactus factorization and crossing suppression as functions of time in Fig. \eqref{fig:xxzfactorcross} (also see Appendix \ref{appendix:xxzothero}). In both the easy-plane and the easy-axis regimes, factorization persists to longer times in the chaotic chains than in the integrable chains. The crossing contribution is also typically at least one order of magnitude smaller in the chaotic cases. Overall, these results show that the integrable spin chains can satisfy part of the full-ETH cumulant structure, including the reconstruction of the OTOC from $k^{\text{ETH}}_2(t)$ and $k^{\text{ETH}}_4(t)$, while still differing from chaotic systems in their finite-size scaling, late-time dynamics, and time-domain factorization properties. We remark that the operator $\hat{O}_{1}$ is related to the spin transport and  $\hat{O}_{2}$ is related to the energy transport. The two transport behave different in the integrable regimes. While spin transport is dependent on $\Delta$, energy transport is independent of $\Delta$. This might possibly result in the different behaviors as observed in Fig. \eqref{fig:xxzfullethballistic} (e,f) and \eqref{fig:xxzfullethdiffusive} (c,d), although we do not establish it conclusively.


\indent

\section{Outlook and Discussions} 
We have tested the full ETH conditions in integrable spin chains by studying the OTOC in two paradigmatic models, the Ising chain and the XXZ chain, using full diagonalization. We observe that the integrable systems can satisfy  a part of the full-ETH conditions. In particular, we find that the relation
$\text{OTOC}(t)\simeq2[k^{\text{ETH}}_{2}(t)]^{2}+k^{\text{ETH}}_{4}(t)$
holds for the integrable models considered here, similar to the chaotic systems studies previously 
\cite{Pappalardi:2023nsj}. This implies that this relation is not a diagnostic of integrable or chaotic dynamics. The distinction instead appears in the scaling behavior of the crossing contributions or quantities like $\braket{r}$ which quantify factorization. In the Ising chain, the crossing contribution remains sub-leading and decreases with system size, but its decay is slower in the integrable case than in the chaotic case. Similarly, cactus factorization holds over a shorter time window in the integrable chain, while it persists to later times in the chaotic chain. These differences are also reflected in the late-time OTOC dynamics: the integrable Ising chain shows persistent oscillations, whereas the chaotic case decays more rapidly.

For the XXZ chain, the conclusions depend on the operator. For $\hat{O}_{1}$, the results are qualitatively similar to the Ising case in both the easy-plane and the easy-axis regimes: the OTOC-cumulant relation holds, while the finite-size scaling and time-domain factorization distinguish integrable and chaotic dynamics. For $\hat{O}_{2}$, however, the scaling of $|\braket{r}-1|$ is similar in the integrable and chaotic cases over the accessible system sizes. This indicates that, for this observable, the cactus-factorization correction in the locally perturbed chaotic XXZ chain is not visibly suppressed as expected from full ETH. The crossing contribution still distinguishes the two regimes, but the behavior of $|\braket{r}-1|$ suggests that some integrable-like features persist for certain local operators. Overall, our results show that integrable spin chains may obey a full-ETH-like cumulant relation for selected observables, but they differ from chaotic systems in the scaling of the corrections and in the time range over which the full-ETH conditions hold. The validity of the OTOC-cumulant relation is thus not a diagnostic of the underlying dynamics and might continue to hold for integrable systems as well.
Several other questions remain open. First, larger system sizes are needed to determine whether the apparent scaling observed here persists asymptotically. Second, the operator dependence found in the locally perturbed XXZ chain, especially the behavior of $\hat{O}_{2}$, requires further clarification. Third, the present work only deals with the fourth order correlation functions and it is possible that the relations analogous to $\text{OTOC}(t)\simeq 2[k^{\text{ETH}}_{2}(t)]^{2}+k^{\text{ETH}}_{4}(t)$ for higher point functions may not hold in integrable systems while continuing to hold for chaotic systems.  Finally, our comparison between the easy-plane and the easy-axis regimes for the XXZ model suggests that the observed scaling behavior is not strongly controlled by the spin-transport properties, but this conclusion should be tested more systematically in future.

\section{Acknowledgments}
The author would like to thank Masaki Tezuka and Viktor Jahnke for careful reading of the draft and for many useful suggestions and comments. 
A part of numerical computations were performed at the computational facilities of the Yukawa Institute for Theoretical Physics. The author also thanks the Yukawa Institute for Theoretical Physics at Kyoto University, and  Viktor Jahnke for his talk there, where this work was initiated during the YITP-I-25-01 on ``Black Hole, Quantum Chaos and Quantum Information". The work is supported by the grant JST CREST (Grant No. JPMJCR24I2).

\section{Data availability} The data that support the findings of this article are not publicly available. The data are available from the authors upon reasonable request.
\balance
\bibliography{references}

@article{PhysRevLett.106.050405,
  title = {Strong and Weak Thermalization of Infinite Nonintegrable Quantum Systems},
  author = {Ba\~nuls, M. C. and Cirac, J. I. and Hastings, M. B.},
  journal = {Phys. Rev. Lett.},
  volume = {106},
  issue = {5},
  pages = {050405},
  numpages = {4},
  year = {2011},
  month = {Feb},
  publisher = {American Physical Society},
  doi = {10.1103/PhysRevLett.106.050405},
  url = {https://link.aps.org/doi/10.1103/PhysRevLett.106.050405}
}

@article{PhysRevLett.111.127205,
  title = {Ballistic Spreading of Entanglement in a Diffusive Nonintegrable System},
  author = {Kim, Hyungwon and Huse, David A.},
  journal = {Phys. Rev. Lett.},
  volume = {111},
  issue = {12},
  pages = {127205},
  numpages = {5},
  year = {2013},
  month = {Sep},
  publisher = {American Physical Society},
  doi = {10.1103/PhysRevLett.111.127205},
  url = {https://link.aps.org/doi/10.1103/PhysRevLett.111.127205}
}

@article{PhysRevE.91.062128,
  title = {Thermalization of entanglement},
  author = {Zhang, Liangsheng and Kim, Hyungwon and Huse, David A.},
  journal = {Phys. Rev. E},
  volume = {91},
  issue = {6},
  pages = {062128},
  numpages = {6},
  year = {2015},
  month = {Jun},
  publisher = {American Physical Society},
  doi = {10.1103/PhysRevE.91.062128},
  url = {https://link.aps.org/doi/10.1103/PhysRevE.91.062128}
}

@article{Roberts:2014isa,
    author = "Roberts, Daniel A. and Stanford, Douglas and Susskind, Leonard",
    title = "{Localized shocks}",
    eprint = "1409.8180",
    archivePrefix = "arXiv",
    primaryClass = "hep-th",
    reportNumber = "MIT-CTP-4594, SU-ITP-14-20",
    doi = "10.1007/JHEP03(2015)051",
    journal = "JHEP",
    volume = "03",
    pages = "051",
    year = "2015"
}

@article{PhysRevB.101.174313,
  title = {Lyapunov growth in quantum spin chains},
  author = {Craps, Ben and De Clerck, Marine and Janssens, Djunes and Luyten, Vincent and Rabideau, Charles},
  journal = {Phys. Rev. B},
  volume = {101},
  issue = {17},
  pages = {174313},
  numpages = {17},
  year = {2020},
  month = {May},
  publisher = {American Physical Society},
  doi = {10.1103/PhysRevB.101.174313},
  url = {https://link.aps.org/doi/10.1103/PhysRevB.101.174313}
}

@article{Rodriguez-Nieva:2023err,
    author = "Rodriguez-Nieva, Joaquin F. and Jonay, Cheryne and Khemani, Vedika",
    title = "{Quantifying Quantum Chaos through Microcanonical Distributions of Entanglement}",
    eprint = "2305.11940",
    archivePrefix = "arXiv",
    primaryClass = "cond-mat.stat-mech",
    doi = "10.1103/PhysRevX.14.031014",
    journal = "Phys. Rev. X",
    volume = "14",
    number = "3",
    pages = "031014",
    year = "2024"
}

@article{Santos_2004,
   title={Integrability of a disordered Heisenberg spin-1/2 chain},
   volume={37},
   ISSN={1361-6447},
   url={http://dx.doi.org/10.1088/0305-4470/37/17/004},
   DOI={10.1088/0305-4470/37/17/004},
   number={17},
   journal={Journal of Physics A: Mathematical and General},
   publisher={IOP Publishing},
   author={Santos, L F},
   year={2004},
   month=apr, pages={4723–4729} }

@article{Alcaraz:1987uk,
    author = "Alcaraz, F. C. and Barber, M. N. and Batchelor, M. T. and Baxter, R. J. and Quispel, G. R. W.",
    title = "{Surface Exponents of the Quantum XXZ, Ashkin-Teller and Potts Models}",
    reportNumber = "ANU-RR-7-1987",
    doi = "10.1088/0305-4470/20/18/038",
    journal = "J. Phys. A",
    volume = "20",
    pages = "6397",
    year = "1987"
}

@article{Pappalardi:2023nsj,
    author = "Pappalardi, Silvia and Fritzsch, Felix and Prosen, Toma{\v{z}}",
    title = "{Full Eigenstate Thermalization via Free Cumulants in Quantum Lattice Systems}",
    eprint = "2303.00713",
    archivePrefix = "arXiv",
    primaryClass = "cond-mat.stat-mech",
    doi = "10.1103/PhysRevLett.134.140404",
    journal = "Phys. Rev. Lett.",
    volume = "134",
    number = "14",
    pages = "140404",
    year = "2025"
}

@misc{noncross,
  note = "\emph{Noncrossing} partition can be understood in following intuitive manner: Consider two blocks (parts) $\mathcal{B}_{1}$ and $\mathcal{B}_{2}$ of a set $\mathcal{S}$. If $a_{1}, b_{1} \in \mathcal{B}_{1}$ (digrammatically an arch with endpoints at $a_{1}$ and $b_{1}$) and $a_{2}, b_{2} \in \mathcal{B}_{2}$ (digrammatically an arch with endpoints at $a_{2}$ and $b_{2}$), then noncrossing partition implies the ordering is such that, the two blocks do not cross (diagrammatically the two arches should not have a crossing). Ordering like $a_{1}a_{2}b_{1}b_{2}$ are crossing while orderings like $a_{1}b_{1}a_{2}b_{2}$ or $a_{1}a_{2}b_{2}b_{1}$ are noncrossing. "
}

@article{Pappalardi:2022aaz,
    author = "Pappalardi, Silvia and Foini, Laura and Kurchan, Jorge",
    title = "{Eigenstate Thermalization Hypothesis and Free Probability}",
    eprint = "2204.11679",
    archivePrefix = "arXiv",
    primaryClass = "cond-mat.stat-mech",
    doi = "10.1103/PhysRevLett.129.170603",
    journal = "Phys. Rev. Lett.",
    volume = "129",
    number = "17",
    pages = "170603",
    year = "2022"
}

@article{Srednicki:1994mfb,
    author = "Srednicki, Mark",
    title = "{Chaos and Quantum Thermalization}",
    eprint = "cond-mat/9403051",
    archivePrefix = "arXiv",
    doi = "10.1103/PhysRevE.50.888",
    journal = "Phys. Rev. E",
    volume = "50",
    month = "3",
    year = "1994"
}

@article{Deutsch:1991msp,
    author = "Deutsch, J. M.",
    title = "{Quantum statistical mechanics in a closed system}",
    doi = "10.1103/PhysRevA.43.2046",
    journal = "Phys. Rev. A",
    volume = "43",
    number = "4",
    pages = "2046",
    year = "1991"
}

@article{Rigol:2007juv,
    author = "Rigol, Marcos and Dunjko, Vanja and Olshanii, Maxim",
    title = "{Thermalization and its mechanism for generic isolated quantum systems}",
    eprint = "0708.1324",
    archivePrefix = "arXiv",
    primaryClass = "cond-mat.stat-mech",
    doi = "10.1038/nature06838",
    journal = "Nature",
    volume = "452",
    number = "7189",
    pages = "854--858",
    year = "2008"
}

@article{Foini:2018sdb,
    author = "Foini, Laura and Kurchan, Jorge",
    title = "{Eigenstate thermalization hypothesis and out of time order correlators}",
    eprint = "1803.10658",
    archivePrefix = "arXiv",
    primaryClass = "cond-mat.stat-mech",
    doi = "10.1103/PhysRevE.99.042139",
    journal = "Phys. Rev. E",
    volume = "99",
    number = "4",
    pages = "042139",
    year = "2019"
}

@article{PhysRevB.98.235128,
  title = {High-temperature coherent transport in the {XXZ} chain in the presence of an impurity},
  author = {Brenes, Marlon and Mascarenhas, Eduardo and Rigol, Marcos and Goold, John},
  eprint = "1810.03640",
    archivePrefix = "arXiv",
    primaryClass = "cond-mat",
  journal = {Phys. Rev. B},
  volume = {98},
  issue = {23},
  pages = {235128},
  numpages = {14},
  year = {2018},
  month = {Dec},
  publisher = {American Physical Society},
  doi = {10.1103/PhysRevB.98.235128},
  url = {https://link.aps.org/doi/10.1103/PhysRevB.98.235128}
}

@article{PhysRevE.84.016206,
  title = {Domain wall dynamics in integrable and chaotic spin-1$/$2 chains},
  author = {Santos, Lea F. and Mitra, Aditi},
  journal = {Phys. Rev. E},
  volume = {84},
  issue = {1},
  pages = {016206},
  numpages = {8},
  year = {2011},
  month = {Jul},
  publisher = {American Physical Society},
  doi = {10.1103/PhysRevE.84.016206},
  url = {https://link.aps.org/doi/10.1103/PhysRevE.84.016206}
}

@article{PhysRevE.89.062110,
  title = {Local quenches with global effects in interacting quantum systems},
  author = {Torres-Herrera, E. J. and Santos, Lea F.},
  journal = {Phys. Rev. E},
  volume = {89},
  issue = {6},
  pages = {062110},
  numpages = {10},
  year = {2014},
  month = {Jun},
  publisher = {American Physical Society},
  doi = {10.1103/PhysRevE.89.062110},
  url = {https://link.aps.org/doi/10.1103/PhysRevE.89.062110}
}

@article{Torres_Herrera_2015,
   title={Relaxation and thermalization of isolated many-body quantum systems},
   volume={T165},
   ISSN={1402-4896},
   url={http://dx.doi.org/10.1088/0031-8949/2015/T165/014018},
   DOI={10.1088/0031-8949/2015/t165/014018},
   journal={Physica Scripta},
   publisher={IOP Publishing},
   author={Torres-Herrera, E J and Kollmar, Davida and Santos, Lea F},
   year={2015},
   month=oct, pages={014018} }

@article{PhysRevB.80.125118,
  title = {Incoherent transport induced by a single static impurity in a Heisenberg chain},
  author = {Bari\ifmmode \check{s}\else \v{s}\fi{}i\ifmmode \acute{c}\else \'{c}\fi{}, O. S. and Prelov\ifmmode \check{s}\else \v{s}\fi{}ek, P. and Metavitsiadis, A. and Zotos, X.},
  journal = {Phys. Rev. B},
  volume = {80},
  issue = {12},
  pages = {125118},
  numpages = {4},
  year = {2009},
  month = {Sep},
  publisher = {American Physical Society},
  doi = {10.1103/PhysRevB.80.125118},
  url = {https://link.aps.org/doi/10.1103/PhysRevB.80.125118}
}

@article{PhysRevLett.125.070605,
  title = {Eigenstate Thermalization in a Locally Perturbed Integrable System},
  author = {Brenes, Marlon and LeBlond, Tyler and Goold, John and Rigol, Marcos},
  journal = {Phys. Rev. Lett.},
  volume = {125},
  issue = {7},
  pages = {070605},
  numpages = {6},
  year = {2020},
  month = {Aug},
  publisher = {American Physical Society},
  doi = {10.1103/PhysRevLett.125.070605},
  url = {https://link.aps.org/doi/10.1103/PhysRevLett.125.070605}
}

@article{Deutsch:2025pmk,
    author = {Deutsch, J. M.},
    title = {A closed quantum system giving ergodicity},
    journal={},
    eprint = {2502.15947},
    archivePrefix = {arXiv},
    primaryClass = {quant-ph},
    month={2},
    year = {2025}
}

@article{PhysRevE.60.3949,
  title = {Ergodic properties of a generic nonintegrable quantum many-body system in the thermodynamic limit},
  author = {Prosen, Toma\check{z}},
  journal = {Phys. Rev. E},
  volume = {60},
  issue = {4},
  pages = {3949--3968},
  numpages = {0},
  year = {1999},
  month = {Oct},
  publisher = {American Physical Society},
  doi = {10.1103/PhysRevE.60.3949},
  url = {https://link.aps.org/doi/10.1103/PhysRevE.60.3949}
}

@article{PhysRevLett.105.250401,
  title = {Effect of Rare Fluctuations on the Thermalization of Isolated Quantum Systems},
  author = {Biroli, Giulio and Kollath, Corinna and L\"auchli, Andreas M.},
  journal = {Phys. Rev. Lett.},
  volume = {105},
  issue = {25},
  pages = {250401},
  numpages = {4},
  year = {2010},
  month = {Dec},
  publisher = {American Physical Society},
  doi = {10.1103/PhysRevLett.105.250401},
  url = {https://link.aps.org/doi/10.1103/PhysRevLett.105.250401}
}

@article{RevModPhys.83.863,
  title = {Colloquium: Nonequilibrium dynamics of closed interacting quantum systems},
  author = {Polkovnikov, Anatoli and Sengupta, Krishnendu and Silva, Alessandro and Vengalattore, Mukund},
  journal = {Rev. Mod. Phys.},
  volume = {83},
  issue = {3},
  pages = {863--883},
  numpages = {0},
  year = {2011},
  month = {Aug},
  publisher = {American Physical Society},
  doi = {10.1103/RevModPhys.83.863},
  url = {https://link.aps.org/doi/10.1103/RevModPhys.83.863}
}

@article{PhysRevE.87.012125,
  title = {Finite-size scaling analysis of the eigenstate thermalization hypothesis in a one-dimensional interacting Bose gas},
  author = {Ikeda, Tatsuhiko N. and Watanabe, Yu and Ueda, Masahito},
  journal = {Phys. Rev. E},
  volume = {87},
  issue = {1},
  pages = {012125},
  numpages = {5},
  year = {2013},
  month = {Jan},
  publisher = {American Physical Society},
  doi = {10.1103/PhysRevE.87.012125},
  url = {https://link.aps.org/doi/10.1103/PhysRevE.87.012125}
}

@article{DAlessio:2015qtq,
    author = "D'Alessio, Luca and Kafri, Yariv and Polkovnikov, Anatoli and Rigol, Marcos",
    title = "{From quantum chaos and eigenstate thermalization to statistical mechanics and thermodynamics}",
    eprint = "1509.06411",
    archivePrefix = "arXiv",
    primaryClass = "cond-mat.stat-mech",
    doi = "10.1080/00018732.2016.1198134",
    journal = "Adv. Phys.",
    volume = "65",
    number = "3",
    pages = "239--362",
    year = "2016"
}

@article{PhysRevLett.117.170404,
  title = {Anomalous Thermalization in Ergodic Systems},
  author = {Luitz, David J. and Bar Lev, Yevgeny},
  journal = {Phys. Rev. Lett.},
  volume = {117},
  issue = {17},
  pages = {170404},
  numpages = {5},
  year = {2016},
  month = {Oct},
  publisher = {American Physical Society},
  doi = {10.1103/PhysRevLett.117.170404},
  url = {https://link.aps.org/doi/10.1103/PhysRevLett.117.170404}
}

@article{Vidmar_2016,
   title={Generalized Gibbs ensemble in integrable lattice models},
   volume={2016},
   ISSN={1742-5468},
   url={http://dx.doi.org/10.1088/1742-5468/2016/06/064007},
   DOI={10.1088/1742-5468/2016/06/064007},
   number={6},
   journal={Journal of Statistical Mechanics: Theory and Experiment},
   publisher={IOP Publishing},
   author={Vidmar, Lev and Rigol, Marcos},
   year={2016},
   month=jun, pages={064007} }

@article{PhysRevLett.113.117202,
  title = {Quenching the Anisotropic Heisenberg Chain: Exact Solution and Generalized Gibbs Ensemble Predictions},
  author = {Wouters, B. and De Nardis, J. and Brockmann, M. and Fioretto, D. and Rigol, M. and Caux, J.-S.},
  journal = {Phys. Rev. Lett.},
  volume = {113},
  issue = {11},
  pages = {117202},
  numpages = {6},
  year = {2014},
  month = {Sep},
  publisher = {American Physical Society},
  doi = {10.1103/PhysRevLett.113.117202},
  url = {https://link.aps.org/doi/10.1103/PhysRevLett.113.117202}
}

@article{PhysRevE.87.012118,
  title = {Eigenstate thermalization within isolated spin-chain systems},
  author = {Steinigeweg, R. and Herbrych, J. and Prelov\ifmmode \check{s}\else \v{s}\fi{}ek, P.},
  journal = {Phys. Rev. E},
  volume = {87},
  issue = {1},
  pages = {012118},
  numpages = {5},
  year = {2013},
  month = {Jan},
  publisher = {American Physical Society},
  doi = {10.1103/PhysRevE.87.012118},
  url = {https://link.aps.org/doi/10.1103/PhysRevE.87.012118}
}

@article{PhysRevB.93.134201,
  title = {Long tail distributions near the many-body localization transition},
  author = {Luitz, David J.},
  journal = {Phys. Rev. B},
  volume = {93},
  issue = {13},
  pages = {134201},
  numpages = {10},
  year = {2016},
  month = {Apr},
  publisher = {American Physical Society},
  doi = {10.1103/PhysRevB.93.134201},
  url = {https://link.aps.org/doi/10.1103/PhysRevB.93.134201}
}

@article{PhysRevB.91.155123,
  title = {Eigenstate thermalization hypothesis and integrability in quantum spin chains},
  author = {Alba, Vincenzo},
  journal = {Phys. Rev. B},
  volume = {91},
  issue = {15},
  pages = {155123},
  numpages = {7},
  year = {2015},
  month = {Apr},
  publisher = {American Physical Society},
  doi = {10.1103/PhysRevB.91.155123},
  url = {https://link.aps.org/doi/10.1103/PhysRevB.91.155123}
}

@article{PhysRevE.91.012144,
  title = {Off-diagonal matrix elements of local operators in many-body quantum systems},
  author = {Beugeling, Wouter and Moessner, Roderich and Haque, Masudul},
  journal = {Phys. Rev. E},
  volume = {91},
  issue = {1},
  pages = {012144},
  numpages = {10},
  year = {2015},
  month = {Jan},
  publisher = {American Physical Society},
  doi = {10.1103/PhysRevE.91.012144},
  url = {https://link.aps.org/doi/10.1103/PhysRevE.91.012144}
}

@article{PhysRevE.102.062113,
  title = {Eigenstate thermalization for observables that break Hamiltonian symmetries and its counterpart in interacting integrable systems},
  author = {LeBlond, Tyler and Rigol, Marcos},
  journal = {Phys. Rev. E},
  volume = {102},
  issue = {6},
  pages = {062113},
  numpages = {15},
  year = {2020},
  month = {Dec},
  publisher = {American Physical Society},
  doi = {10.1103/PhysRevE.102.062113},
  url = {https://link.aps.org/doi/10.1103/PhysRevE.102.062113}
}

@article{Rottoli:2025pzr,
    author = "Rottoli, Federico and Alba, Vincenzo",
    title = "{Eigenstate Thermalization Hypothesis (ETH) for off-diagonal matrix elements in integrable spin chains}",
    eprint = "2505.23602",
    archivePrefix = "arXiv",
    primaryClass = "cond-mat.stat-mech",
    month = "5",
    year = "2025",
    journal=""
}

@article{PhysRevLett.106.217206,
  title = {Open {XXZ} Spin Chain: Nonequilibrium Steady State and a Strict Bound on Ballistic Transport},
  author = {Prosen, Toma\check{z}},
  journal = {Phys. Rev. Lett.},
  volume = {106},
  issue = {21},
  pages = {217206},
  numpages = {4},
  year = {2011},
  month = {May},
  publisher = {American Physical Society},
  doi = {10.1103/PhysRevLett.106.217206},
  url = {https://link.aps.org/doi/10.1103/PhysRevLett.106.217206}
}

@article{RevModPhys.93.025003,
  title = {Finite-temperature transport in one-dimensional quantum lattice models},
  author = {Bertini, B. and Heidrich-Meisner, F. and Karrasch, C. and Prosen, T. and Steinigeweg, R. and \ifmmode \check{Z}\else \v{Z}\fi{}nidari\ifmmode \check{c}\else \v{c}\fi{}, M.},
  journal = {Rev. Mod. Phys.},
  volume = {93},
  issue = {2},
  pages = {025003},
  numpages = {71},
  year = {2021},
  month = {May},
  publisher = {American Physical Society},
  doi = {10.1103/RevModPhys.93.025003},
  url = {https://link.aps.org/doi/10.1103/RevModPhys.93.025003}
}

@article{Maldacena:2015waa,
    author = "Maldacena, Juan and Shenker, Stephen H. and Stanford, Douglas",
    title = "{A bound on chaos}",
    eprint = "1503.01409",
    archivePrefix = "arXiv",
    primaryClass = "hep-th",
    doi = "10.1007/JHEP08(2016)106",
    journal = "JHEP",
    volume = "08",
    pages = "106",
    year = "2016"
}

@article{PROSEN1994115,
title = {Statistical Properties of Matrix Elements in a Hamilton System Between Integrability and Chaos},
journal = {Annals of Physics},
volume = {235},
number = {1},
pages = {115-164},
year = {1994},
issn = {0003-4916},
doi = {https://doi.org/10.1006/aphy.1994.1093},
url = {https://www.sciencedirect.com/science/article/pii/S0003491684710931},
author = {T. Prosen}
}

@article{PhysRevLett.123.260601,
  title = {Eigenstate Thermalization and Rotational Invariance in Ergodic Quantum Systems},
  author = {Foini, Laura and Kurchan, Jorge},
  journal = {Phys. Rev. Lett.},
  volume = {123},
  issue = {26},
  pages = {260601},
  numpages = {4},
  year = {2019},
  month = {Dec},
  publisher = {American Physical Society},
  doi = {10.1103/PhysRevLett.123.260601},
  url = {https://link.aps.org/doi/10.1103/PhysRevLett.123.260601}
}

@article{PhysRevLett.122.220601,
  title = {Eigenstate Correlations, Thermalization, and the Butterfly Effect},
  author = {Chan, Amos and De Luca, Andrea and Chalker, J. T.},
  journal = {Phys. Rev. Lett.},
  volume = {122},
  issue = {22},
  pages = {220601},
  numpages = {5},
  year = {2019},
  month = {Jun},
  publisher = {American Physical Society},
  doi = {10.1103/PhysRevLett.122.220601},
  url = {https://link.aps.org/doi/10.1103/PhysRevLett.122.220601}
}

@article{PhysRevLett.123.230606,
  title = {Bounds on Chaos from the Eigenstate Thermalization Hypothesis},
  author = {Murthy, Chaitanya and Srednicki, Mark},
  journal = {Phys. Rev. Lett.},
  volume = {123},
  issue = {23},
  pages = {230606},
  numpages = {5},
  year = {2019},
  month = {Dec},
  publisher = {American Physical Society},
  doi = {10.1103/PhysRevLett.123.230606},
  url = {https://link.aps.org/doi/10.1103/PhysRevLett.123.230606}
}

@article{PhysRevE.104.034120,
  title = {Out-of-time-order correlations and the fine structure of eigenstate thermalization},
  author = {Brenes, Marlon and Pappalardi, Silvia and Mitchison, Mark T. and Goold, John and Silva, Alessandro},
  journal = {Phys. Rev. E},
  volume = {104},
  issue = {3},
  pages = {034120},
  numpages = {13},
  year = {2021},
  month = {Sep},
  publisher = {American Physical Society},
  doi = {10.1103/PhysRevE.104.034120},
  url = {https://link.aps.org/doi/10.1103/PhysRevE.104.034120}
}

@article{PhysRevE.102.042127,
  title = {Eigenstate thermalization hypothesis beyond standard indicators: Emergence of random-matrix behavior at small frequencies},
  author = {Richter, Jonas and Dymarsky, Anatoly and Steinigeweg, Robin and Gemmer, Jochen},
  journal = {Phys. Rev. E},
  volume = {102},
  issue = {4},
  pages = {042127},
  numpages = {12},
  year = {2020},
  month = {Oct},
  publisher = {American Physical Society},
  doi = {10.1103/PhysRevE.102.042127},
  url = {https://link.aps.org/doi/10.1103/PhysRevE.102.042127}
}

@article{larkin1969quasiclassical,
  title={Quasiclassical method in the theory of superconductivity},
  author={Larkin, Anatoly I and Ovchinnikov, Yu N},
  journal={Sov Phys JETP},
  volume={28},
  number={6},
  pages={1200--1205},
  year={1969}
}

@article{Hosur:2015ylk,
    author = "Hosur, Pavan and Qi, Xiao-Liang and Roberts, Daniel A. and Yoshida, Beni",
    title = "{Chaos in quantum channels}",
    eprint = "1511.04021",
    archivePrefix = "arXiv",
    primaryClass = "hep-th",
    reportNumber = "MIT-CTP-4733",
    doi = "10.1007/JHEP02(2016)004",
    journal = "JHEP",
    volume = "02",
    pages = "004",
    year = "2016"
}

@article{Hashimoto:2017oit,
    author = "Hashimoto, Koji and Murata, Keiju and Yoshii, Ryosuke",
    title = "{Out-of-time-order correlators in quantum mechanics}",
    eprint = "1703.09435",
    archivePrefix = "arXiv",
    primaryClass = "hep-th",
    doi = "10.1007/JHEP10(2017)138",
    journal = "JHEP",
    volume = "10",
    pages = "138",
    year = "2017"
}

@article{Roberts:2016hpo,
    author = "Roberts, Daniel A. and Yoshida, Beni",
    title = "{Chaos and complexity by design}",
    eprint = "1610.04903",
    archivePrefix = "arXiv",
    primaryClass = "quant-ph",
    doi = "10.1007/JHEP04(2017)121",
    journal = "JHEP",
    volume = "04",
    pages = "121",
    year = "2017"
}

@ARTICLE{Garcia-Mata:2023,
AUTHOR = {Garc\'ia-Mata, I.  and Jalabert, R. A. and Wisniacki, D. Ariel},
TITLE   = {{O}ut-of-time-order correlations and quantum chaos},
YEAR    = {2023},
JOURNAL = {Scholarpedia},
VOLUME  = {18},
NUMBER  = {4},
PAGES   = {55237},
DOI     = {10.4249/scholarpedia.55237}
}

@article{PhysRevLett.111.057203,
  title = {Families of Quasilocal Conservation Laws and Quantum Spin Transport},
  author = {Prosen, Toma\check{z} and Ilievski, Enej},
  journal = {Phys. Rev. Lett.},
  volume = {111},
  issue = {5},
  pages = {057203},
  numpages = {5},
  year = {2013},
  month = {Aug},
  publisher = {American Physical Society},
  doi = {10.1103/PhysRevLett.111.057203},
  url = {https://link.aps.org/doi/10.1103/PhysRevLett.111.057203}
}

@article{Jindal:2024zcg,
    author = "Jindal, Siddharth and Hosur, Pavan",
    title = "{Generalized free cumulants for quantum chaotic systems}",
    eprint = "2401.13829",
    archivePrefix = "arXiv",
    primaryClass = "cond-mat.stat-mech",
    doi = "10.1007/JHEP09(2024)066",
    journal = "JHEP",
    volume = "09",
    pages = "066",
    year = "2024"
}

@article{PhysRevX.13.011045,
  title = {Coherent Fluctuations in Noisy Mesoscopic Systems, the Open Quantum SSEP, and Free Probability},
  author = {Hruza, Ludwig and Bernard, Denis},
  journal = {Phys. Rev. X},
  volume = {13},
  issue = {1},
  pages = {011045},
  numpages = {28},
  year = {2023},
  month = {Mar},
  publisher = {American Physical Society},
  doi = {10.1103/PhysRevX.13.011045},
  url = {https://link.aps.org/doi/10.1103/PhysRevX.13.011045}
}

@inbook{danfreeprob,
author = {Dan Voiculescu},
title = {Free Noncommutative Random Variables, Random Matrices and the ${II}_1$ Factors of Free Groups},
booktitle = {Quantum Probability and Related Topics},
chapter = {},
pages = {473-487},
doi = {10.1142/9789814360203_0031},
URL = {https://www.worldscientific.com/doi/abs/10.1142/9789814360203_0031}
}

@article{PhysRevB.111.014311,
  title = {Free independence and the noncrossing partition lattice in dual-unitary quantum circuits},
  author = {Chen, Hyaline Junhe and Kudler-Flam, Jonah},
  journal = {Phys. Rev. B},
  volume = {111},
  issue = {1},
  pages = {014311},
  numpages = {18},
  year = {2025},
  month = {Jan},
  publisher = {American Physical Society},
  doi = {10.1103/PhysRevB.111.014311},
  url = {https://link.aps.org/doi/10.1103/PhysRevB.111.014311}
}

@article{Jahnke:2025exd,
    author = "Jahnke, Viktor and Nandy, Pratik and Pal, Kuntal and Camargo, Hugo A. and Kim, Keun-Young",
    title = "{Free Probability approach to spectral and operator statistics in Rosenzweig-Porter random matrix ensembles}",
    eprint = "2506.04520",
    archivePrefix = "arXiv",
    primaryClass = "hep-th",
    reportNumber = "RIKEN-iTHEMS-Report-25",
    month = "6",
    year = "2025",
    journal=" "
}

@article{Camargo:2025zxr,
    author = "Camargo, Hugo A. and Fu, Yichao and Jahnke, Viktor and Pal, Kuntal and Kim, Keun-Young",
    title = "{Quantum Signatures of Chaos from Free Probability}",
    eprint = "2503.20338",
    archivePrefix = "arXiv",
    primaryClass = "hep-th",
    month = "3",
    year = "2025",
    journal=""
}

@article{Fritzsch:2025arx,
    author = "Fritzsch, Felix and Claeys, Pieter W.",
    title = "{Free Probability in a Minimal Quantum Circuit Model}",
    eprint = "2506.11197",
    archivePrefix = "arXiv",
    primaryClass = "quant-ph",
    month = "6",
    year = "2025",
    journal=""
}

@article{Dowling:2025cxr,
    author = "Dowling, Neil and De Nardis, Jacopo and Heinrich, Markus and Turkeshi, Xhek and Pappalardi, Silvia",
    title = "{Free Independence and Unitary Design from Random Matrix Product Unitaries}",
    eprint = "2508.00051",
    archivePrefix = "arXiv",
    primaryClass = "quant-ph",
    month = "7",
    year = "2025",
    journal=""
}

@article{Fritzsch:2025ban,
    author = "Fritzsch, Felix and Alves, Gabriel O. and Rampp, Michael A. and Claeys, Pieter W.",
    title = "{Free Cumulants and Full Eigenstate Thermalization from Boundary Scrambling}",
    eprint = "2509.08060",
    archivePrefix = "arXiv",
    primaryClass = "quant-ph",
    month = "9",
    year = "2025",
    journal=""
}

@Inbook{Speicher2003,
author="Speicher, R.",
editor="Vershik, Anatoly M.
and Yakubovich, Yuri",
title="Free Probability Theory and Random Matrices",
bookTitle="Asymptotic Combinatorics with Applications to Mathematical Physics: A European Mathematical Summer School held at the Euler Institute, St. Petersburg, Russia July 9--20, 2001",
year="2003",
publisher="Springer Berlin Heidelberg",
address="Berlin, Heidelberg",
pages="53--73",
isbn="978-3-540-44890-7",
doi="10.1007/3-540-44890-X_3",
url="https://doi.org/10.1007/3-540-44890-X_3"
}

@article{speicher2017free,
  title={Free probability theory: and its avatars in representation theory, random matrices, and operator algebras; also featuring: non-commutative distributions},
  author={Speicher, Roland},
  journal={Jahresbericht der Deutschen Mathematiker-Vereinigung},
  volume={119},
  number={1},
  pages={3--30},
  year={2017},
  publisher={Springer}
}

@book{mingo2017free,
  title={Free probability and random matrices},
  author={Mingo, James A and Speicher, Roland},
  volume={35},
  year={2017},
  publisher={Springer}
}

@book{nica2006lectures,
  title={Lectures on the combinatorics of free probability},
  author={Nica, Alexandru and Speicher, Roland},
  volume={13},
  year={2006},
  publisher={Cambridge University Press}
}

@article{PhysRevX.14.031048,
  title = {Statistics of Matrix Elements of Local Operators in Integrable Models},
  author = {Essler, F. H. L. and de Klerk, A. J. J. M.},
  journal = {Phys. Rev. X},
  volume = {14},
  issue = {3},
  pages = {031048},
  numpages = {29},
  year = {2024},
  month = {Sep},
  publisher = {American Physical Society},
  doi = {10.1103/PhysRevX.14.031048},
  url = {https://link.aps.org/doi/10.1103/PhysRevX.14.031048}
}

@article{PhysRevB.111.054303,
  title = {Microcanonical free cumulants in lattice systems},
  author = {Fritzsch, Felix and Prosen, Toma{\v{z}} and Pappalardi, Silvia},
  journal = {Phys. Rev. B},
  volume = {111},
  issue = {5},
  pages = {054303},
  numpages = {16},
  year = {2025},
  month = {Feb},
  publisher = {American Physical Society},
  doi = {10.1103/PhysRevB.111.054303},
  url = {https://link.aps.org/doi/10.1103/PhysRevB.111.054303}
}
\clearpage
\onecolumngrid
\appendix
\section{Ising model: Chaotic case}\label{appendix:isingchao}
We first show results for Ising model, with open boundary conditions, and chaotic parameters: $h_{x}= -1.05, h_{z}= 0.5$, for various sizes $L$. Ising model with chaotic parameters and periodic boundary conditions are already considered in \cite{Pappalardi:2023nsj}. The results here merely complement those results. 
\begin{figure}[h]
    \centering
    \includegraphics[width= 0.9\linewidth]{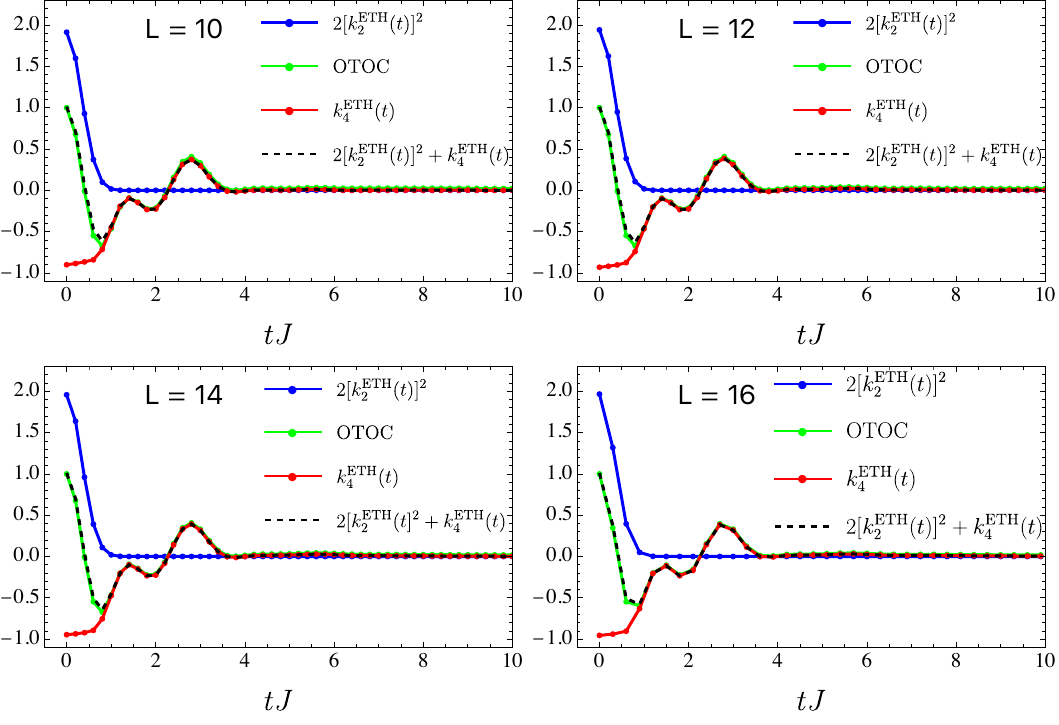}
    \caption{The time evolution OTOC, $2[k^{\mathrm{ETH}}_{2}(t)]^{2}$ and $k^{\mathrm{ETH}}_{4}(t)$ for Ising model with chaotic parameters and system sizes $L =10, 12, 14, 16$. Also shown is the sum $2[k^{\mathrm{ETH}}_{2}(t)]^{2}+k^{\mathrm{ETH}}_{4}(t)$.}
\label{fig:isingfullethchaosupp}
\end{figure}

We observe from Fig. \eqref{fig:isingfullethchaosupp} that the OTOC quickly goes to zero , unlike the integrable case where we observe oscillation or saturation to finite constant value for accessible times. The fourth order ETH free cumulant, $k^{\mathrm{ETH}}_{4}(t)$ is found to govern the late time behavior of OTOC.

\clearpage

\section{Ising model: Results for other local operators}\label{appendix:othero}
We now numerically study the full ETH condition for other local operators. Specifically, we will test the following conditions \cite{Foini:2018sdb, Pappalardi:2022aaz}

(i) The crossing contribution are exponentially subleading. For instance (relevant contribution at fourth order)
\begin{equation}\label{seqn:fulleth1supp}
    \text{cross}(t)= \frac{1}{D} \sum_{i \neq j}e^{i t \omega_{ij}}|O_{ij}|^4= \mathcal{O}(D^{-a}),\, a>0. 
\end{equation}

(ii) The non-crossing diagram (also called cactus diagram) factorize. For instance (relevant contribution at fourth order)
\begin{equation}\label{seqn:fulleth2supp}
\text{cac}(t_{1},t_{2})= \frac{1}{D}\sum_{i \neq j \neq k} e^{i \omega_{ij}t_{1}+ i \omega_{jk} t_{2}} |O_{ij}|^{2} |O_{jk}|^{2} \approx k^{\mathrm{ETH}}_{2}(t_{1}) k^{\mathrm{ETH}}_{2}(t_{2})
\end{equation}
for generic times $t_{1}$ and $t_{2}$ and $D$ denotes the Hilbert space dimension.

\begin{figure}[H]
    \centering
    \includegraphics[width= .9\linewidth]{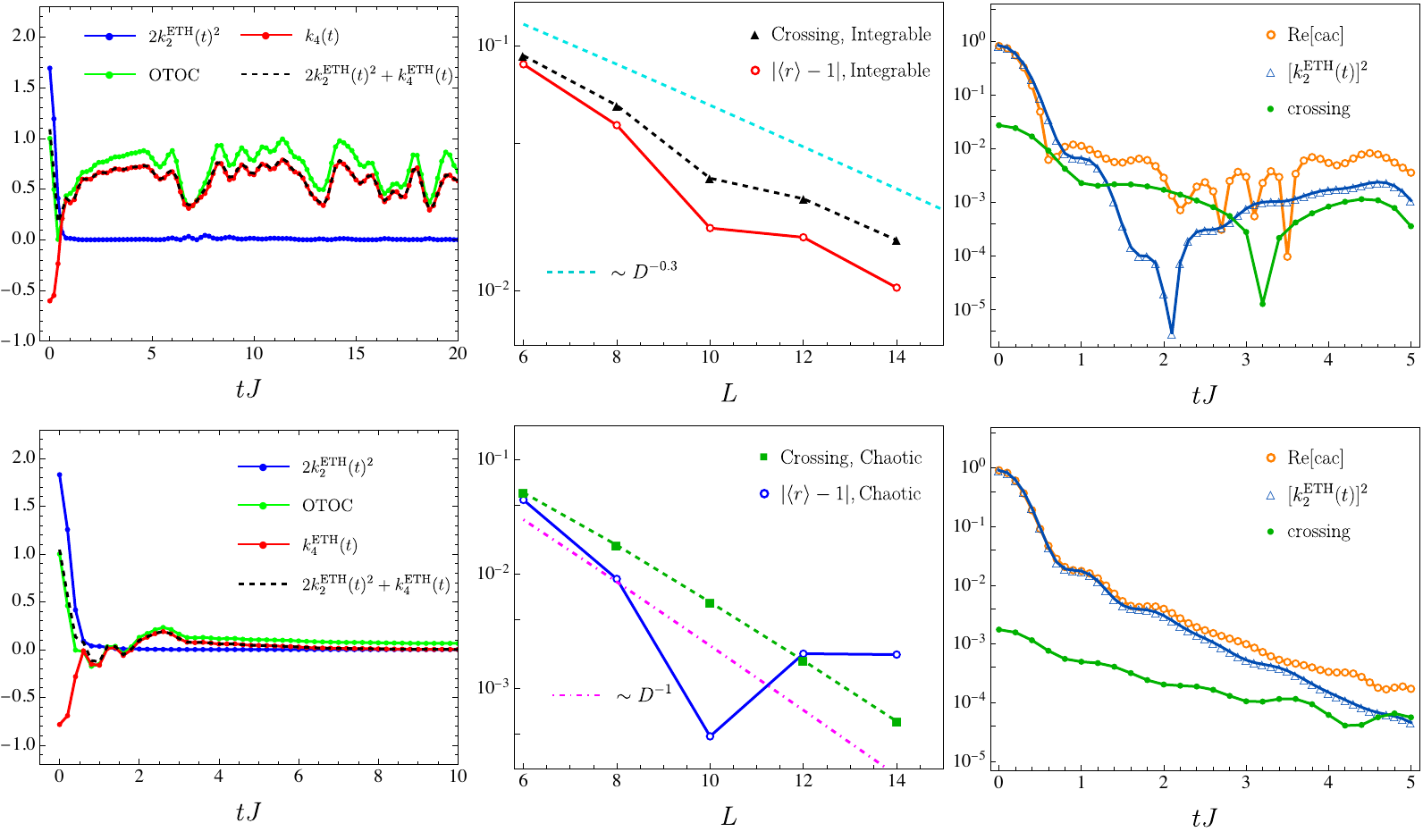}
    \caption{(First row) Left: The time evolution OTOC, $2[k^{\mathrm{ETH}}_{2}(t)]^{2}$ and $k^{\mathrm{ETH}}_{4}(t)$. Also shown is the sum of the two free cumulant contributions against OTOC for integrable Ising model with $L = 12$.  Middle: Variation of $|\langle r \rangle -1|$ and crossing contribution for integrable Ising model. Also shown are the $D^{-0.3}$ scaling line. Right: Crossing contribution as a function of time, Eq. \eqref{seqn:fulleth1supp}, and factorization of cactus diagram in time domain as given by Eq. \eqref{seqn:fulleth2supp} for integrable case.
    Second row: Same as the first row but for chaotic case.}
    \label{fig:isingfullethx}
\end{figure}

\begin{figure}[H]
    \centering
    \includegraphics[width=0.9 \linewidth]{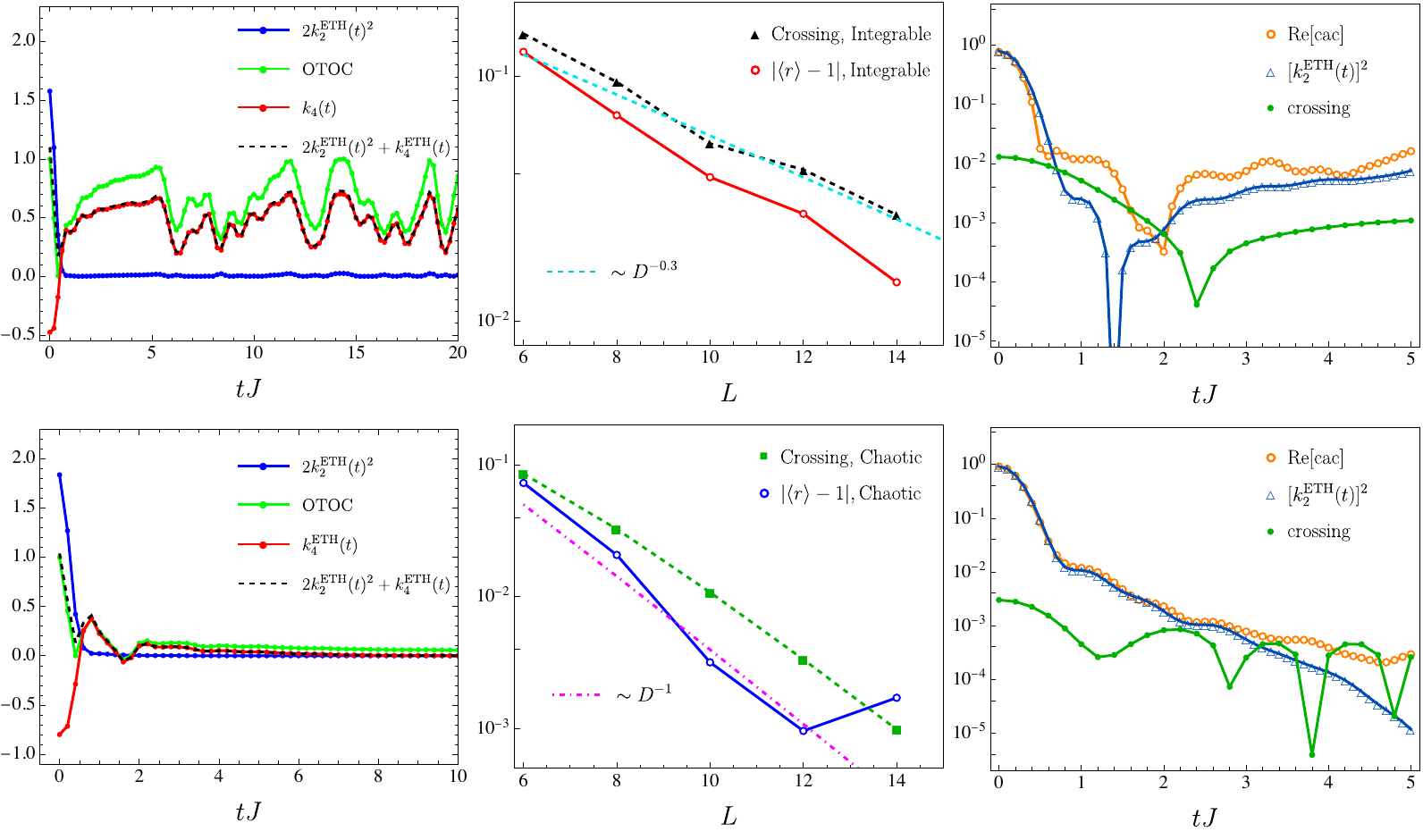}
    \caption{Same as Fig. \eqref{fig:isingfullethx} but for operator $\hat{O}= \sigma^{z}_{L/2}\sigma^{z}_{L/2+1}$.}
    \label{fig:isingfullethzz}
\end{figure}

First we show the results for the Ising model with open boundary conditions and consider the operators: $\sigma^{x}_{L/2}$ and $\sigma^{z}_{L/2}\sigma^{z}_{L/2+1}$. We consider both the integrable and the chaotic parameters of the model and contrast their finding in Fig. \eqref{fig:isingfullethx} and Fig. \eqref{fig:isingfullethzz} for operators $\sigma^{x}_{L/2}$ and $\sigma^{z}_{L/2}\sigma^{z}_{L/2+1}$ respectively. 

\section{XXZ model}\label{appendix:xxzothero}
\subsection{Easy-plane regime}
We now show supporting results for the XXZ model. The results presented are for following two operators
\begin{equation}
  \hat{O}_{1} = \sigma_{L/4}^{z} , \quad \hat{O}_{2} = \sigma_{3L/4}^{x}\sigma_{3L/4+1}^{x} +\sigma_{3L/4}^{y}\sigma_{3L/4+1}^{y}
\end{equation} 
We perform full diagonalization of spin chains with $L=16$ spins and anisotropy parameter $\Delta= .55$ corresponding to the easy-plane regime.

\begin{figure}[htbp]
    \centering
    \includegraphics[width= 0.81\linewidth]{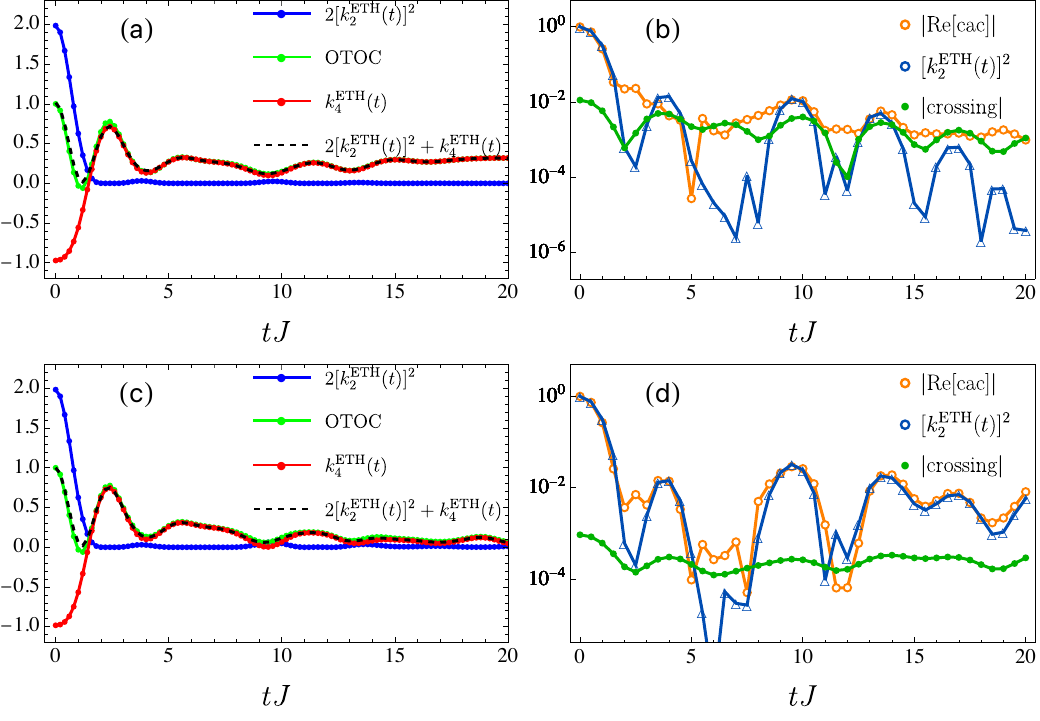}
    \caption{Evolution of OTOC, $2[k^{\mathrm{ETH}}_{2}(t)]^{2}$ and $k^{\mathrm{ETH}}_{4}(t)$. Also shown is the sum of the two free cumulant contributions, against OTOC for (a) integrable XXZ model (c) chaotic XXZ model, with $L = 16$. Factorization of cactus diagram as given by Eq.\eqref{seqn:fulleth2supp} and the crossing contribution for (b) integrable XXZ model($d=0$) and (d) chaotic XXZ model ($d=1$). All results are shown for operator $\hat{O}_{1}$. }
    \label{fig:xxzfullethz}
\end{figure}

\begin{figure}[htbp]
    \centering
    \includegraphics[width= .81\linewidth]{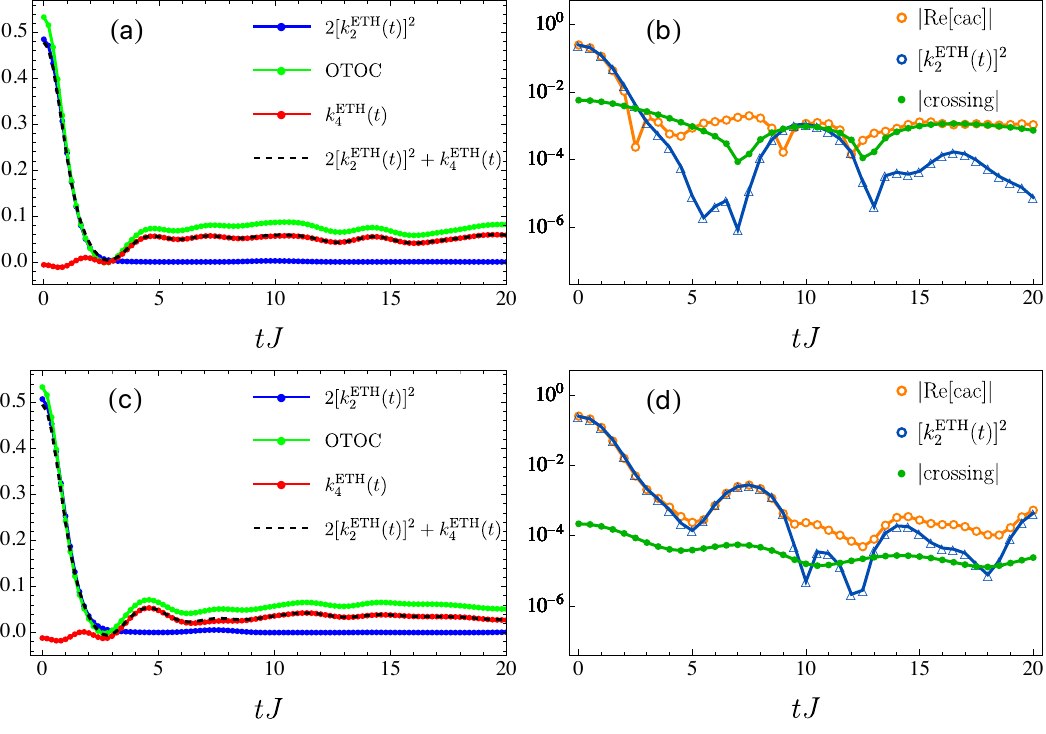}
    \caption{Same as Fig. \eqref{fig:xxzfullethz} but for operator $\hat{O}_{2}$.}
    \label{fig:xxzfullethxy}
\end{figure}

In Fig. \eqref{fig:xxzfullethz} (a) and (c) results are shown for the evolution of OTOC, $2[k^{\mathrm{ETH}}_{2}(t)]^{2}$ and $k^{\mathrm{ETH}}_{4}(t)$, using operator $\hat{O}_{1}$, for integrable and chaotic cases respectively. In Fig. \eqref{fig:xxzfullethz} (b) and (d) we test the factorization condition, Eq. \eqref{seqn:fulleth2supp} for the integrable and chaotic cases respectively. In Fig. \eqref{fig:xxzfullethxy} the results (with same labellings) are shown for operator $\hat{O}_{2}$.
\clearpage
\subsection{Easy-axis regime}
\begin{figure}[H]
    \centering
    \includegraphics[width= .9\linewidth]{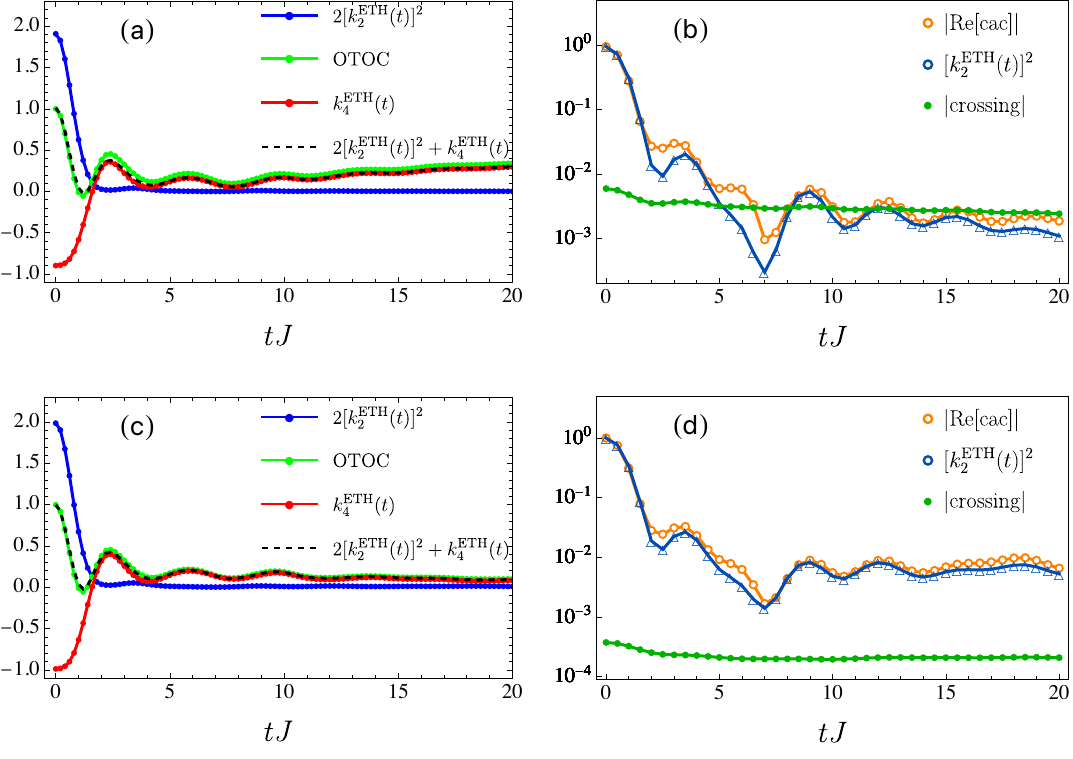}
    \caption{Time evolution of OTOC, $2[k^{\mathrm{ETH}}_{2}(t)]^{2}$ and $k^{\mathrm{ETH}}_{4}(t)$. Also shown is the sum of the two free cumulant contributions against OTOC for (a) integrable XXZ model (c) chaotic XXZ model, with $L = 18$. Factorization of cactus diagram as given by Eq.\eqref{seqn:fulleth2supp} and the crossing contribution for (b) integrable XXZ model($d=0$) and (d) chaotic XXZ model ($d=1$). All results are shown for operator $\hat{O}_{1}$ and anisotropy parameter $\Delta= 1.1$ corresponding to diffusive spin transport. }\label{fig:xxzfullethxydiff18}
\end{figure}

 In this section, we consider the XXZ model with $L=18$ spins with anisotropy parameter $\Delta= 1.1$ corresponding to diffusive spin transport. In Fig. \eqref{fig:xxzfullethxydiff18} (a) and (c) results are shown for the evolution of OTOC, $2[k^{\mathrm{ETH}}_{2}(t)]^{2}$ and $k^{\mathrm{ETH}}_{4}(t)$, using operator $\hat{O}_{1}$, for integrable and chaotic cases respectively. In Fig. \eqref{fig:xxzfullethxydiff18} (b) and (d) we test the factorization condition, Eq. \eqref{seqn:fulleth2supp} for the integrable and chaotic cases respectively.
\end{document}